\documentstyle[graphics,preprint,aps]{revtex}

\tightenlines

\begin{document}
\draft
\title{Resonances in one and two rows of triangular Josephson junction cells}
\author{P.~Caputo\cite{addr1}, M.~V.~Fistul\cite{addr2}, and A.~V.~Ustinov} \address{Physikalisches Institut III, Universit\"at Erlangen-N\"urnberg \\ Erwin-Rommel-Str. 1, D-91058 Erlangen, Germany}

\date{\today}

%\wideabs{ %REVTeX 3.1 feature

\maketitle

\begin{abstract}

We present an experimental and analytical study of resonances in the
current-voltage characteristics of single and double row triangular Josephson arrays.
The magnetic field dependences of the voltage positions of the resonances have been measured for various cell inductances
and junction critical currents. In double row arrays, we have observed a peculiar resonance, whose voltage position decreases with magnetic field.
We derive the spectrum of linear
electromagnetic waves propagating in the arrays.
In the double row array, the spectrum consists of many branches that differ by the behavior of the Josephson junctions transverse to the bias current direction.
The measured magnetic field dependence of the resonance voltages is mapped to the linear mode spectrum and good agreement between experiments and model is found.

\end{abstract}


%}
%end of WideAbs

\section{Introduction}
The driven Josephson junction ladders and arrays have recently attracted a wide interest due to many fascinating phenomena observed in such systems. They show vortex propagation\cite{ParArrays,Ustinov:PRB95}, various current-voltage resonances\cite{caputo,Barahona:PRB97}, nonlinear dynamic localized modes\cite{Binder:PRL00,Trias:PRL00}, and may also be practically useful as high-frequency oscillators\cite{Yukon,Caputo:ASC00}.

A peculiarity of quasi-two-dimensional arrays, {\it i.e.} Josephson ladders, and two-dimensional arrays is the presence of Josephson junctions in both the longitudinal and transverse directions to the bias current [see sketches in Figs.\,\ref{sketches}(a)--(c)]. In these systems, the  Josephson phase dynamics is more complex than in the rather well studied case of one-dimensional parallel arrays\cite{ParArrays,Ustinov:PRB95}, due to the higher degree of freedom appearing from the presence of the junctions transverse to the bias.
For example, the spectrum $\omega(q)$ of the linear electromagnetic waves (EWs) propagating in the ladders [Fig.\,\ref{sketches}(a)] contains {\it two} branches \cite{caputo,Barahona:PRB97,Yukon}. Similar behavior has been extensively studied in vertical stacks of long Josephson junctions [see Ref.\,\onlinecite{Ustinov:frascati98} and references therein], where the inductive coupling between the junctions leads to the splitting of the dispersion relation. The splitting has been found also in coupled one-dimensional parallel arrays\cite{Duwel:JAP96}.
As we turn to the case of two-dimensional arrays, the number of branches in the spectrum $\omega(q)$ increases. This feature has been recently confirmed by a systematic numerical study on dynamical states in underdamped arrays\cite{Rotoli:ASC00}.

It is well known that, in the presence of magnetic field, the nonlinear interaction between the
Josephson current wave and the excited EWs leads to resonant steps in the current-voltage characteristics of extended Josephson systems \cite{ParArrays,Brill,Barone}. The measured magnetic field
dependence of the voltages of the resonant steps can be mapped to the spectrum $\omega(q)$ by the relationship:
\begin{equation}\label{ResVoltage}
 V_{\rm res}\,=\,\frac{\hbar \omega(q)}{2e},~~~~q\,=\,f~ \quad, 
\end{equation}
where $f\,=\,\Phi_{\rm ext}/\Phi_0$ is the magnetic flux threading the cell
normalized to the magnetic flux quantum, and it is referred to as frustration parameter.
The spectrum $\omega(q)$ for the Josephson ladder with four junctions per elementary cell [Fig.\,\ref{sketches}(a)] has been calculated in Ref.\,\onlinecite{caputo} in the linear approximation, and a good agreement between the measured magnetic field dependence of the step voltages and Eqs.\,(\ref{ResVoltage}) was found.
This system has a remarkable property that at $f\,=\,0.5$
an {\em "out\,-\,of\,-\,phase"} state of the currents is formed.
As the amplitude of the resonant step reaches a maximum, the direction of the mesh currents flowing in adjacent cells alternate from
one cell to another, and the so-called "{\em checkerboard\,\,}" structure of the alternating (ac) currents appears \cite{Barahona:PRB97,Yukon}. The ac currents compensate each other on the horizontal junctions and the total alternating voltage along the ladder is close to zero. This property is an obvious drawback if one wants to use the ladder as an oscillator. To overcome it, Yukon and Lin  proposed \cite{Yukon} to use a {\em triangular} arrangement of Josephson junctions along the ladder [Figs.\,\ref{sketches}(b)--(c)]. In this case, at  $f\,=\,0.5$ an "{\em in-phase\,}" state with all the ac currents flowing in the same direction can be realized for the junctions transverse to the dc bias current.
So far, the detailed study of the EWs dispersion in triangular Josephson ladders and arrays has not been carried out, as well as the study of the dependence of the resonant steps on the magnetic field. Thus, in this paper we present measurements of the resonant steps in the current-voltage ($I$--$V$) characteristics of single and double row triangular Josephson arrays. For both systems, we calculate the spectrum of linear EWs and map it on the magnetic field dependence of the step voltages.
The paper is arranged as follows: in Sec. II our experimental observations are presented. In Sec. III, we introduce the model based on a set of equations, and derive the spectrum of linear EWs.
Finally, Sec. IV contains a discussion of the obtained results and conclusions.

\section{Experiments with triangular Josephson junction ladders and arrays}

In order to investigate the dynamical behavior of triangular arrays in the presence of an external magnetic field, we have chosen linear arrays made of one row and two rows, as shown in Figs.\,\ref{sketches}(b)-(c).
Arrays of various critical current densities and sizes of the elementary cells have been measured.
The arrays were made of
Nb/Al-AlO$_x$/Nb
underdamped Josephson tunnel junctions, arranged in a triangular lattice.
The junction area is designed to be 9\,$\mu$m$^2$. The studied critical current densities are $j_{c}\,\approx\,50$\,A/cm$^2$ and $j_{c}\,\approx\,1000$\,A/cm$^2$.
The junction capacitance $C$ is about $300\,$pF and $450\,$pF, respectively.
The bias current $I_b$ is uniformly injected in each node of the array
via on-chip resistors and extracted as shown by arrows in the sketches.
The voltage is measured in the direction along the bias, across each individual row. The measurements have been performed using the
acquisition software \mbox{Ref.\,\onlinecite{GoldExi}}.
Experiments have been performed in the presence of a magnetic field
applied perpendicular to the cell plane. Following the standard notation, we express the magnetic field in terms of the frustration $f$.
The self-inductance of one cell can be roughly estimated \cite{Jaycox} as
\begin{equation}
  L = 1.25\,\mu_0\sqrt{A} \quad,
\label{inductance}
\end{equation}
\noindent where $\mu_0$ is the free space permeability, and $A$ is the cell hole size.
The value of $L$ is needed for calculating the parameter $\beta_L\,=\,2\pi L I_c/\Phi_0$, where
$I_c$ is the junction critical current.
The value given by Eq.\,\ref{inductance} underestimates the actual inductance of the cell,
as
it is an asymptotic value valid only when the width of the
superconducting electrodes forming the cell is larger than the hole size \cite{Jaycox}. On the contrary, our arrays are in the opposite limit.
We have compared the calculated value (Eq.\,\ref{inductance}) with
the value $\widetilde L$ obtained from the magnetic field dependence of the critical current of a single cell with two-junctions, {\it i.e.} a SQUID. The ratio of the minimum to the maximum critical currents of the SQUID reflects
the value of its $\beta_L$ parameter\cite{Barone} and, therefore, the value of the cell inductance\cite{squid}.
The relation between the two values was found to be $L\,\approx\,0.7\,\widetilde L$.
However, in order to make systematic comparison of all studied
arrays, the $\beta_L$ values reported here refer to the cell
inductance estimated from Eq.\,\ref{inductance}.
To get various values of $\beta_L$, we have used samples either with
different critical current densities and same cell size, or have varied the cell
size $A$ from $126\,\mu$m$^2$ to
$240\,\mu$m$^2$ for similar arrays located on one chip.
Thus, the parameter $\beta_L$ varied between $0.5$ to $5$, at the temperature $T\,=\,$4.2\,K.
The investigated arrays and their parameters are summarized in Table\,I.

The measured $I$--$V$ curves of the triangular ladders show well defined resonances, grouped in two different voltage regions.
Fig.\,\ref{IVs}(a) shows an enlargement of the $I$--$V$ curve of a ten cell row with $\beta_L\,\approx\,0.5$ at $f\,=\,0.3$. Here $S_1$ and $S_2$ denote the upper and lower voltage resonances, respectively.
The steps of type $S_1$ and $S_2$ are stable in approximately the same range of frustration. Changes of $f$ induce a periodic modulation of the voltages $V_{1,2}$ of the steps. In particular,
$S_1$ and $S_2$ approach the maximum voltage at $f\,=\,0.5$ (and other half-integer values), and tend to the minimum value of voltage at integer $f$.
Figure\,\ref{IVs}(b) shows the $V_{1,2}(f)$ dependences in the range
$0 \leq f \leq 1$. The step $\it{S_2}$ has a typical resonant behavior. It clearly shows distinctly different voltages (underlined by the dashed lines of Fig.\,\ref{IVs}(b)) that correspond to the minima of the differential resistance on the step. However, we have noticed that the resonant regime of the step $\it{S_2}$ is realized only in the case of low $\beta_L$, {\it i.e.} for small discreteness. In arrays with either larger $I_c$ or larger $L$ (both these quantities lead to an increase of $\beta_L$)
the resonant behavior disappears and the step $\it{S_2}$ changes continuously with $f$.

The voltage and the differential resistance of the step $\it{S_1}$ have almost a continuous dependence on $f$ for all the investigated $\beta_L$ parameters.
In general, the differential resistance of the step $\it{S_1}$ is higher than that of $\it{S_2}$.
This feature is enhanced in the samples with large $\beta_L$,
as Fig.\,\ref{smooth} shows for the case of a single row array of 12 cells and $\beta_L\,\approx\,5$. At a constant bias current chosen on the step $\it{S_1}$, by changing $f$ we could continuously tune the voltage across the array, along the periodic pattern shown in the figure inset.

The maximum voltage of the steps $S_1$ and $S_2$, {\it i.e.} the voltage value approached at $f\,=\,0.5$,
depends on the cell inductance. We have compared the maximum voltages of three triangular ladders (samples \#3, \#4, \#5) designed with different cell sizes ($A\,=\,126\,\mu{\rm m}^2$, $180\,\mu{\rm m}^2$ and $240\,\mu{\rm m}^ 2$, respectively) and same junction area, and fabricated with the same critical current density ($j_{c}\,\approx\,50\,{\rm A/cm}^2$).
%These arrays consist of one row of 10 cells.
The data reported in Fig.\,\ref{inductances} show a tendency for the maximum voltages of both $S_1$ and $S_2$ to decrease as the $\beta_L$ (cell size) increases.
This effect seems
to be more pronounced for the step $S_2$ than for the step $S_1$.
Increasing temperature also leads to a reduction of the maximum step voltages, similarly to the results reported in Ref.\onlinecite{Barahona:PRB97}.

In experiments with two row arrays [Fig.\,\ref{sketches}(c)], we have used a common bias current for both rows, and have independently measured the voltage across each of them.
Figure\,\ref{IVs_two_rows} shows the typical $I$--$V$ characteristics of an array made of $12{\rm(cells)}\times{\rm2(rows)}$, sample\,\#6.
Both rows (denoted as A and B) have very similar $I$--$V$ curves. At the same critical current, the rows switch simultaneously to the finite voltage state, with the same voltage. Similar to the case of single row arrays, in the presence of frustration we distinguish the steps $\it{S_1}$ and $\it{S_2}$. In Fig.\,\ref{IVs_two_rows} these steps are plotted at $f\,=\,0.4$. No hysteresis is observed in this case. The steps are found to be periodically modulated by the magnetic field, in a similar way as it is described above for the single row array. Moreover, their maximum voltages approached at $f\,=\,0.5$ are similar to that observed in the single row array with the same discreteness parameter (sample\,\#2), {\it i.e.} with the same cell size ($A\,=\,160\,\,{\rm \mu m}^2$) and critical current density ($j_c\,=\,1050\,{\rm A/cm^2}$).
The new feature observed in the two row arrays is the presence of a third step, marked as $\it{S_3}$ in Fig.\,\ref{IVs_two_rows}. The peculiarity of this resonance is that it is stable at frustration values corresponding to an integer number of fluxons per cell ($f\,=\,0,\pm1,...$).
At these values of $f$, an increase of $I_b$ above the array critical current induces both rows to jump simultaneously to the state $\it{S_3}$.
Further increasing of $I_b$ causes the transition from $\it{S_3}$ to the McCumber branch. The $I$--$V$ curve is hysteretic and a decrease of $I_b$ along $\it{S_3}$ eventually reaches an instability point with a certain retrapping current, at which the array returns to the
zero voltage state. Sometimes, by decreasing $I_b$, at the instability point the two rows split and while one row goes to $V\,=\,0$, the other row undergoes a transition to higher voltages and then to $V\,=\,0$, with a slightly lower retrapping current (this behavior is shown in Fig.\,\ref{IVs_two_rows}).

The dependences of the maximum voltages of the steps
$\it{S_1}$, $\it{S_2}$ and $\it{S_3}$ on $f$ are shown in Fig.\,\ref{dispersion}.
In contrast to $\it{S_1}$ and $\it{S_2}$, the step $\it{S_3}$ moves to lower voltages when approaching $f\,=\,0.5$.
We note that the voltage modulation is more pronounced for the steps $\it{S_1}$ and $\it{S_2}$ than for $\it{S_3}$. Moreover, increasing the temperature above 4.2\,K causes only a slight reduction of the step voltage. Above $f\,\approx \,0.2$ the step $\it{S_3}$ disappears and the steps $\it{S_1}$ and $\it{S_2}$ become stable.

%%%%%%%%%%%%%%%%%%%%%%%%%%%%%%%%%%%%%%%%%%%%%%%%%%%%%%%%%%%%%%%%%%%%

\section{Model and spectrum of linear electromagnetic waves}

The derivation of the equations of motion for a triangular array is done in a way similar to the case of 1D-arrays (2 junctions per cell)\cite{Barahona:PRB97} and ladders (four junctions per cell)\cite{caputo,GriFil:PLA96}.
The junctions are described by the
RCSJ model\,\cite{Barone}. We neglect mutual inductances and consider only the self inductances of the cells.
We denote with $\varphi_{i,j}$ and $\varphi_{i+1,j}$, respectively, the superconducting phase differences (Josephson phases) across the $\em{vertical}$ junctions of the cell $i$ in the row $j$ ($j\,=\,1,2$) and with $\psi_{i,k}$ the phase differences across the $\em{horizontal}$ junction of the cell $i$ in the line $k$ ($k\,=\,1,2,3$). For the one-row case the indexes $j$ and $k$ are unnecessary and, therefore, omitted.
With $\Phi_i$ and $\Phi_{\rm ext}$ we denote the induced and the applied magnetic flux, respectively. First, we derive the equations of motion for the "triangular" Josephson ladder of Fig.\,1(b). We
recall the fluxoid quantization in the cell $i$:

\begin{equation}
  \varphi_{i+1} - \varphi_i + \psi_i \,=\,-\frac{2\pi\Phi_i}{\Phi_0}
  \quad . \label{Eq:1}
\end{equation}
%
%
%where $\Phi_i$ is the induced magnetic flux threading the cell $i$ %and $\Phi_0$ is the magnetic flux quantum.
%Due to non zero cell inductance $L_i$, the induced flux $\Phi_i$ and the applied
%flux $\Phi_{\rm ext}$ are related by
Due to nonzero cell inductance $L$, $\Phi_i$ and $\Phi_{\rm ext}$ are related by
\begin{equation}
  \Phi_i\,=\,\Phi_{\rm ext} + L I_i
  \quad , \label{Eq:2}
\end{equation}
where $I_i$ is the mesh current in the cell $i$.
By making use of Kirchhoff's current law, we get

\begin{equation}
  C \dot V_i^v + \frac{V_i^v}{R} + I_{c} \sin\varphi_i\,=\,I_b - I_i + I_{i-1}
  \quad , \label{rsjH}
\end{equation}
where $V_i^v$ is the voltage across the vertical junction $i$. All junctions have capacitance $C$, resistance $R$ and critical current $I_c$.
Since the current flowing through the horizontal branch is the mesh current $I_i$, for the horizontal junctions we obtain %
\begin{equation}
  C \dot V_i^h + \frac{V_i^h}{R} + I_{c} \sin\psi_i\,=\,I_i
  \quad . \label{rsjV}
\end{equation}
%

%$V_{i,j}^v$ ($V_{i,k}^h$)
%is the voltage across the vertical (horizontal) junction. All junctions have %capacitance $C$, resistance $R$ and critical current $I_{c}$. $I_i$ %is the mesh current in the cell $i$, and %$I_b$ is the bias current.
%
%
Finally, the equations of motion for the vertical and horizontal junctions of the row read as:
\begin{eqnarray}
  \ddot \varphi_i + \alpha \dot \varphi_i + \sin\varphi_i\,&=&\, \nonumber\\
  \,=\,\gamma &+&
\frac{1}{\beta_L} {\left(\varphi_{i+1} - 2\varphi_i + \varphi_{i-1} + \psi_i + \psi_{i-1}\right)}\quad  \nonumber\\
  \ddot \psi_i + \alpha \dot \psi_i + \sin\psi_i\,&=&\, \frac{1}{\beta_L} {\left(\varphi_i - \varphi_{i+1} - \psi_i \right)} - \frac{2\pi f}{\beta_L}
    \quad . \label{Eq:f}
\end{eqnarray}
Here, the time unit is $\omega_{\rm p}^{-1}\,=\,\sqrt{\hbar C/(2e I_c)}$, the inverse plasma frequency. The parameter $\alpha\,=\,1/\sqrt{\beta_{c}}$ determines the damping\cite{Barone} of the junctions and the parameter $\beta_L$ defines the discreteness of the array; $f$ is the frustration defined above, and
$\gamma\,=\,I_{b}/I_c$ is the normalized bias current.

To derive the spectrum of EWs,
we assume a whirling solution along the vertical junctions and oscillations with a small amplitude for the horizontal junctions. Moreover, the phase of the vertical junctions increases from cell to cell due to the presence of frustration.
Thus, the solutions of Eqs.\,\ref{Eq:f} can be written as:
\begin{eqnarray}
\label{PhaseSol}
\varphi_n\,&=\,&\omega t+2\pi f n +\varphi\,e^{i(\omega t+2\pi qn)}  \nonumber\\ \psi_n\,&=\,&\psi\,e^{i(\omega t+2\pi qn)}\quad ,
\end{eqnarray}
where $\omega$ and $q$ are, respectively, the angular frequency and the wave number of the EW in the array. In the limit of small amplitudes $\varphi$ and $\psi$, we obtain
the spectrum of electromagnetic wave propagating along the array. This spectrum consists of {\em two} branches $\omega_{+}(q)$ and $\omega_{-}(q)$ given by:
\begin{equation} \label{Spectrum}
\omega_{\pm}\,=\,\omega_p\sqrt{{\cal F} \pm \sqrt{{\cal F}^2-{\cal G}}} , \end{equation}
where ${\cal F}\,=\,1/2\,+\,1/(2 \beta_L)\,+\, \,(2/\beta_L)\sin^2(\pi q)$, and
${\cal G}\,=\,(4/\beta_L)\sin^2(\pi q)$. The two modes are plotted in Fig.\,\ref{linmodes1}.
This dispersion relation differs from that derived in \mbox{Ref.\,\onlinecite{caputo}} for the case of "square" Josephson ladders [four instead than three small junctions per elementary cell, Fig.\,\ref{sketches}(a)] only by a constant factor.
The horizontal junctions play an essential role in the array dynamics, and lead to two linear resonances in the dispersion relation. If, instead of the horizontal junction, simply a superconducting link is placed, that is the case of well-known 1D parallel array, only {\em one} linear mode exists\cite{ParArrays,Ustinov:PRB95}.

Thus, what should we expect when two rows of cells join together
in a 2D array?
In order to describe the spectrum of linear modes in the two-row array, we use the time dependent Josephson phases of vertical $\varphi_{i,j}(t)$ and $\varphi_{i+1,j}(t)$ and horizontal $\psi_{i,k}(t)$ junctions.
Similar to the one row case, we derive the set of equations by means of the Kirchhoff's current law and the fluxoid quantization.
Assuming a whirling solution along the vertical junctions and oscillations with a small amplitude for the horizontal junctions, we obtain the spectrum of linear modes from the system of seven linear equations:

\begin{eqnarray}
\label{System1}
-\beta_L\omega^2\varphi_{i,1}\,=\,\varphi_{i+1,1}(1+e^{-iq})-2\varphi_{i,1}-\psi_{0,i+1}-\psi_{-1,i}e^{-iq} \quad ,\nonumber \\  
-\beta_L\omega^2 \varphi_{i+1,1}\,=\,\varphi_{i,1}(1+e^{iq})-2\varphi_{i+1,1}+\psi_{-1,i}+\psi_{0,i+1} \quad , \nonumber\\
 -\beta_L\omega^2\varphi_{i,2} \,=\,\varphi_{i+1,2}(1+e^{-iq})-2\varphi_{i,2}+\psi_{0,i+1}+\psi_{1,i}e^{-iq} \quad , \nonumber\\
 -\beta_L\omega^2\varphi_{i+1,2}\,=\,\varphi_{i,2}(1+e^{iq})-2\varphi_{i+1,2}-\psi_{0,i+1}-\psi_{1,i} \quad ,\\  \beta_L(-\omega^2+1)\psi_{1,i}\,=\,(\varphi_{i,2}e^{iq}-\varphi_{i+1,2}-\psi_{1,i}) \quad ,\nonumber\\
 \beta_L(-\omega^2+1) \psi_{-1,i}\,=\,(\varphi_{i,1}e^{iq}-\varphi_{i+1,1}-\psi_{-1,i}) \quad , \nonumber \\
 (-\omega^2+1+\frac{1}{\beta_L})(\psi_{-1,i}+\psi_{1,i})\,=\,  \psi_{0,i+1}\frac{(1-\omega^2)(e^{iq}-1)}{\beta_L\omega^2}\nonumber \quad . \end{eqnarray}
By solving the system of Eqs.\,(\ref{System1}), we obtain {\em seven} branches in the $\omega(q)$ dependence.
Three branches are determined by the equation: %
\begin{eqnarray}\label{3branch:ribbon}
\omega^6-(\frac{5}{\beta_L}+1)\omega^4+(\frac{4}{\beta_L}+\frac{6}{\beta_L^2}  -\frac{4}{\beta_L^2}\cos^2\frac{q}{2})\omega^2 +  \nonumber\\  -\frac{4}{\beta_L^2} \sin^2 \frac{q}{2}\,=\,0 \quad.
\end{eqnarray}
These branches correspond to a {\em ribbon} state \cite{Yukon}, as the
Josephson junctions of the middle row are not active ($\psi_0\,=\,0$).
The lower ribbon branches displays an $\omega(q)$ dependence
which increases as the wave vector increases.
As observed in the experiments, the upper branch displays a different behaviour with respect to the lower branches, {\it i.e.} $\omega$ increases
as the wave vector decreases (alternatively, the voltage increases as the frustration decreases).

The other four solutions of the system of Eqs.\,(\ref{System1}) involve oscillations of the junctions in the middle row and, therefore, correspond to a
{\em checkerboard} state (at $f\,=\,0.5$).
These four branches are determined by the equation: %
\begin{eqnarray}\label{4branch:ribbon}
 [(-\omega^2+\frac{2}{\beta_L})\beta_L^2(1-\omega^2+\frac{1}{\beta_L})-2]\times \nonumber \\
 \times [-2\omega^2+(\frac{2}{\beta_L}-\omega^2)(1-\omega^2)]\, =\,2(1-\omega^2)^2\cos^2\frac{q}{2} \quad.
\end{eqnarray}

As in the ribbon case, the upper branch of the checkerboard state has a different dependence on $f$ with respect to the lower checkerboard branches.
All the linear modes calculated for the double row array are plotted in Fig.\,\ref{linmodes2}. As one can see, the qualitative behaviour of these linear modes is in a good 
agreement with the experimental data reported in Fig.\,\ref{dispersion}. Thus, the linear approximation used for modeling the dynamic behavior of the frustrated arrays is sufficient to quantitatively explain the most essential features observed in the experiments.

\section{Discussion and Conclusions}
In the previous section we derived the spectrum $\omega(q)$ of EWs for triangular Josephson ladders and for double row arrays.
In the Josephson ladder case [Fig.\,\ref{sketches}(b)], we obtain {\em two} branches $\omega_{\pm}(q)$ [Fig.\,\ref{linmodes1}].
The excitation of the EWs that account for these branches
leads to resonant steps in the $I$--$V$ curves [Fig.\,\ref{IVs}]. Although the current amplitude of the resonant steps depends on the array parameters in a rather complicated way, {\em the voltage positions} of the steps can be mapped to the spectrum of
linear EWs by making use of the Eq.(\ref{ResVoltage}). Indeed, we obtain a good agreement between the magnetic
field (frustration $f$) dependence of the step voltages [Fig.\,\ref{IVs}(b)] and the calculated spectrum $\omega_{\pm}$ (Fig.\,\ref{linmodes1}).
The dependence of the maximum voltages on the discreteness parameter $\beta_L$ shown in Fig.\,\ref{inductances}, displays also a
good agreement with the theoretical prediction. Thus, the
limiting operation voltage (i.e. frequency) for each mode can be controlled
by the geometrical inductance and/or by the critical current density.

The calculated spectrum of EWs for the double row triangular Josephson array contains {\em seven} branches (Fig.\,\ref{linmodes2}). Three of them correspond to the
ribbon state, while the others are due to the checkerboard state.
However, the experimental data show that only three branches could be excited in the Josephson arrays and, correspondingly, only three resonances appear in the current-voltage characteristics. The reason of this might be ascribed to internal instabilities of the states.
The observed resonances $S_1$, $S_2$ and $S_3$ (Fig.\,\ref{dispersion}) can be mapped to the spectrum of EWs (Fig.\,\ref{linmodes2}).
Moreover, the observed peculiar resonance ($S_3$) appears
at the small values of frustration $f$, and its voltage {\em decreases} with $f$.
Two branches with the similar behaviour are also found in the spectrum $\omega(q)$ of EWs (Fig.\,\ref{linmodes2}), one corresponding to the checkerboard and the other to the ribbon state. Unfortunately,
measurements of $I$--$V$ curves do not allow to distinguish between these branches and, correspondingly, between the ribbon and checkerboard states.
A distinction can be, in principle, done by detecting radiation from the array due to properly coupled ac voltage.

In conclusion, we have studied the dynamical states of triangular arrays of Josephson junctions in the presence of a magnetic field. As expected, the number of exhibited resonances increases with the number of degrees of freedom of the system. We found that in one-row arrays there are two states, and that in two-row arrays three states are observed. The proposed analytical model allows to obtain the spectrum of linear modes, both for single and double row triangular arrays. The voltage position of the observed resonances is mapped to this spectrum and good agreement
between experiments and theory is found.

{\bf Acknowledgments}

We thank Stanford Yukon and Giacomo Rotoli for valuable and stimulating discussions. We are grateful to Marcus Schuster for his critical reading of the manuscript. Samples were made in part at Research Centre Juelich (FZJ) and in part
at Hypres\cite{Hypres}.
The European Office of
Aerospace Research and Development (EOARD) and the Alexander von Humboldt Stiftung are acknowledged for supporting this work.

\begin{table}[tbp]
 \caption{The Arrays Parameters at $T\,=\,$4.2\,K.}
  \label{table1}
  \center
  \begin{tabular}{lcccccr}
 ${\rm Sample}^a$ & \#1 & \#2 & \#3 & \#4 & \#5 &\#6\\
    \tableline
    ${\rm No.\;of\;rows}$  & 1 & 1 & 1 & 1 & 1 & 2\\
    ${\rm No.\;of\;cells\;per\;row}$  & 10 & 12 & 10 & 10 & 10 & 12\\
    $\rm{Cell\;area}\,(\mu \rm{m}^2)$ & 126 & 160 & 126 & 180 & 240 & 160\\
    $j_c\,(\rm A/cm^2)$ & 30 & 1050 & 50 & 50 & 50 & 1050\\
    $\beta_L$      & 0.5 & 5 & 1.6 & 1.9 & 2.2 & 5\\
\hline
\multicolumn{7}{l}{\footnotesize $^a$All the studied samples have 3 junctions per cell.}
\\
\end{tabular}
\end{table}

%%%%%%%%%%%%%%%%%%%%%Fig.1%%%%%%%%%%%%%%%%%%%%%%%%%%%%%%%%

\begin{figure}[!htbp]
  \centering
  \vspace{0.3cm}
  \caption{
    Sketches of Josephson ladders with (a) four and (b) three 
   Josephson junctions per cell, and of a two-row array (c);
    $\varphi_{i,j}$ and $\varphi_{i+1,j}$ denote the Josephson phases of the vertical junctions of cell $i$ in row $j$; $\psi_{i,k}$ denotes the Josephson phases of the horizontal junction of cell $i$ in line $k$. The magnetic field is applied
perpendicular to the plane of the cells and induces a mesh current $I_i$.
}
\label{sketches}
\end{figure}

%%%%%%%%%%%%%%%%%%%%%Fig.2%%%%%%%%%%%%%%%%%%%%%%%%%%%%%%%%

\begin{figure}[!htbp]
  \centering
  \caption{
   (a) $I$--$V$ characteristic of a one-row array
   at $f\,=\,0.3$; (b) step voltage dependences vs.\,$f$
   of $\it{S_1}$ and $\it{S_2}$ (sample \#1); dashed lines represent
   the well-defined resonances of the step $\it{S_2}$. Here
   $\beta_L\,=\,$0.5 and $T=4.2\,$K.
    }
    \label{IVs}
\end{figure}

%%%%%%%%%%%%%%%%%%%%%Fig.3%%%%%%%%%%%%%%%%%%%%%%%%%%%%%%%%

\begin{figure}[!htbp]
  \centering\leavevmode
  \caption{
    Sequence of $\it{S_1}$ steps recorded while slowly changing $f$ in the range 0--0.5. The single row array has 12 cells and a $\beta_L\,\approx\,5$ (sample \#2). Straight arrows indicate the transition from the step to the McCumber branch. Inset: the voltage across
    the array at the bias point $I_b\,\approx\,32\,\mu$A is smoothly tuned from 0 to $65\,\mu$V as $f$ changes.
    }
    \vspace{0.2cm}
 \label{smooth}
\end{figure}
%

%%%%%%%%%%%%%%%%%%%%%Fig.4%%%%%%%%%%%%%%%%%%%%%%%%%%%%%%%%

%
\begin{figure}[!htbp]
  \label{max_vol}
  \centering\leavevmode
  \caption{
    Filled symbols: Maximum voltage of the steps $\it{S_1}$ and $\it{S_2}$ 
    measured at $f\,=\,0.5$ in arrays having three different values of
    cell inductance $L$ (samples \#3,\#4,\#5). These arrays have 10 cells, $j_c\,\approx\,50{\rm A/cm^2}$ and $\omega_p\,\approx\,7$\,GHz.
    Solid line is the theoretical prediction 
    (Eq.\,9) for the upper resonance, for $q\,=\,0.5$.}
\label{inductances}
\end{figure}
%

%%%%%%%%%%%%%%%%%%%%%Fig.5%%%%%%%%%%%%%%%%%%%%%%%%%%%%%%%%

\begin{figure}[!htbp]
  \centering\leavevmode
  \caption{
    $I$--$V$ characteristics for the two row array showing the steps $\it{S_1}$ and $\it{S_2}$ at $f\,=\,0.4$ (open and filled circles), and the extra step $\it{S_3}$ at $f\,=\,0.1$ (open and filled squares).
    Data refer to the voltages measured independently across the individual rows, noted as A and B. Straight arrows indicate the hysteretic path. Sample \#6, $\beta_L\,=\,5$.
   }
   \label{IVs_two_rows}
\end{figure}

%%%%%%%%%%%%%%%%%%%%%Fig.6%%%%%%%%%%%%%%%%%%%%%%%%%%%%%%%%

\begin{figure}[!htbp]
  \centering\leavevmode
  \caption{
    Measured dependences of the maximum voltages of the steps $\it{S_1}$, $\it{S_2}$ and $\it{S_3}$ as a function of frustration for the two row array (sample\,\#6).
Data refer to the voltage measured across one row (A), another row (B) exhibited simultaneously the same voltage state.
    }
    \label{dispersion}
\end{figure}

%%%%%%%%%%%%%%%%%%%%%Fig.7%%%%%%%%%%%%%%%%%%%%%%%%%%%%%%%%

\begin{figure}[!htb]
\centering
\caption{Linear modes $\omega_+$ and $\omega_-$ in one-row array with three junctions per cell (continuous line) and four junctions per cell (dashed line). $\beta_L\,=\,1$.}
\label{linmodes1}
\end{figure}

%%%%%%%%%%%%%%%%%%%%%Fig.8%%%%%%%%%%%%%%%%%%%%%%%%%%%%%%%%

\begin{figure}[!htb]
\centering
\vspace{0.3cm}
\caption{Calculated spectrum for the linear modes in the two-row array and sketches of the ac mesh current distribution at $f\,=\,0.5$.}
\label{linmodes2}
\end{figure}


\begin{references}
\bibitem{squid} The measured $I_c$ vs. $f$ dependences for the single cell with 2 junctions and the single cell with 3 junctions did not reveal substantial differences, as expected in the large inductance limit.
\end{references}
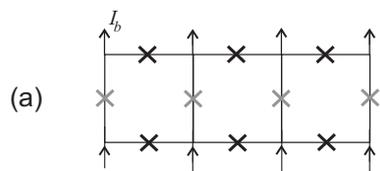
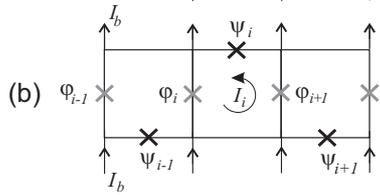
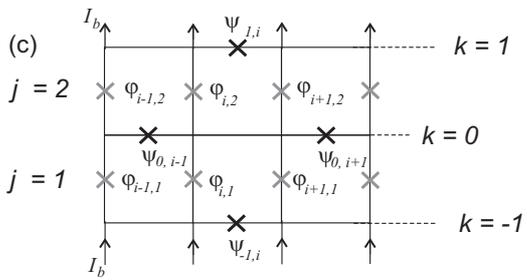
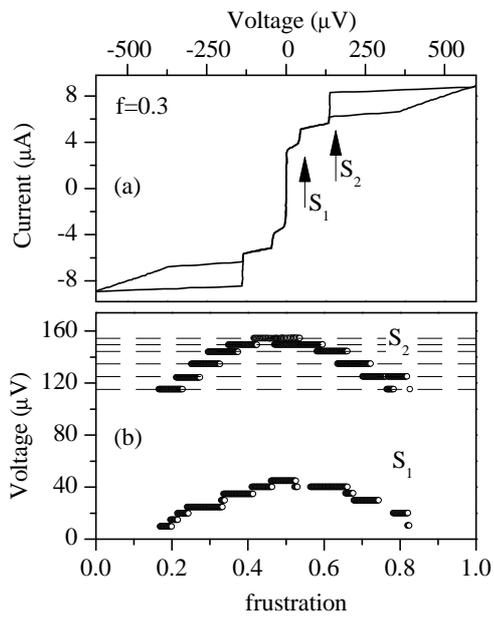
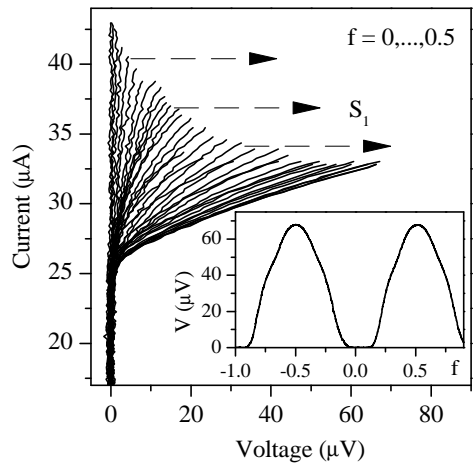
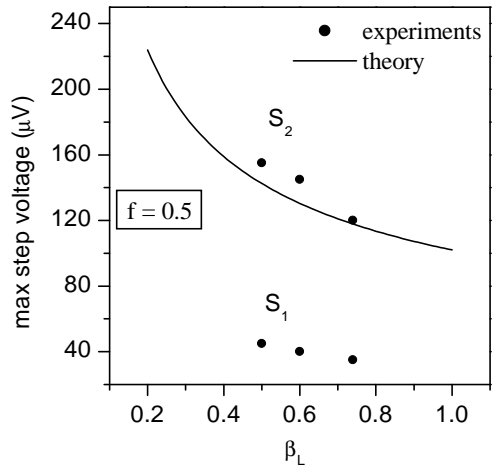
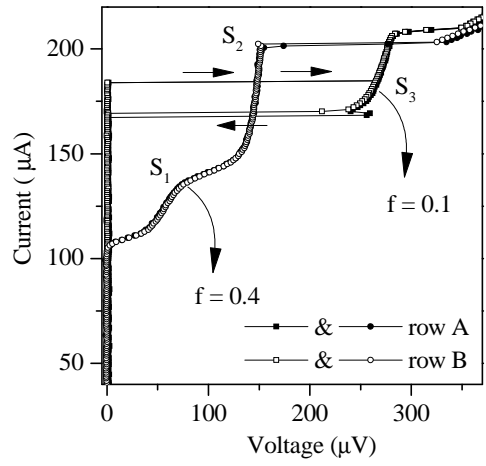
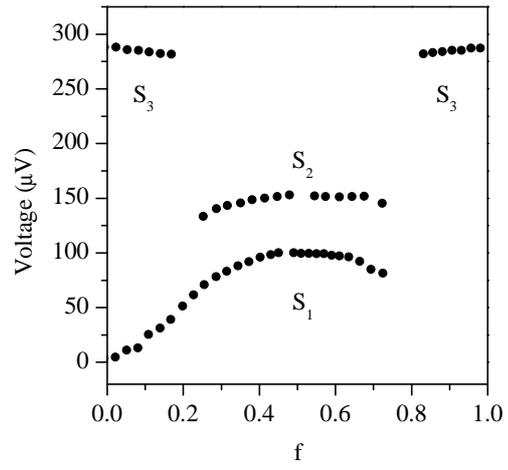
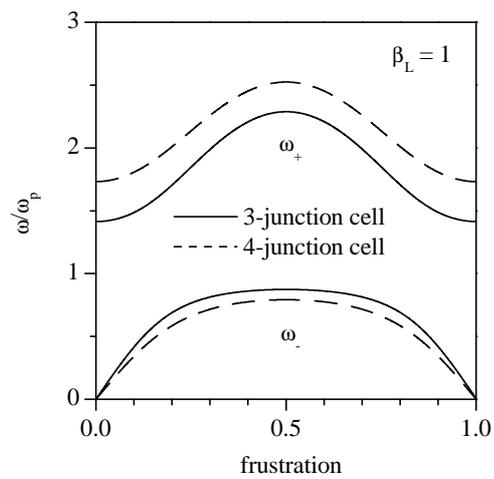
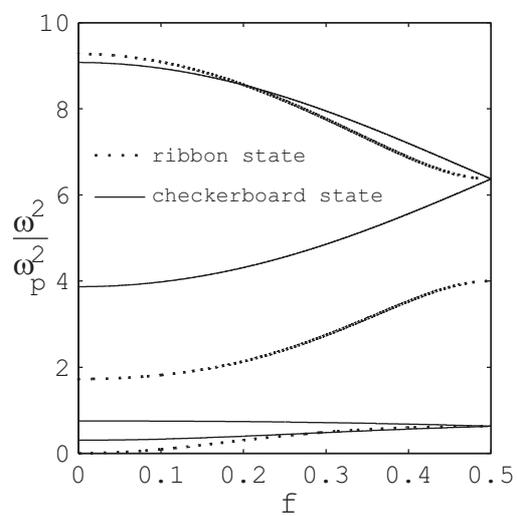
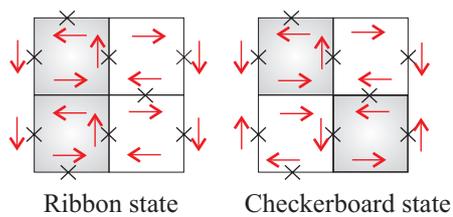
\end{document}